\pdfoutput=1
\documentclass[preprint,12pt]{elsarticle}

\usepackage{booktabs}
\usepackage{multirow}
\usepackage{siunitx}

\usepackage{pdfpages} 
\usepackage{psfrag}

\usepackage{fontenc}
\usepackage[english]{babel}
\usepackage{amsmath}
\usepackage{amsthm}

\usepackage{amscd}
\usepackage{amscd}
\usepackage{amsbsy}
\usepackage{amsfonts}
\usepackage{mathrsfs} 
\usepackage[section]{placeins}

\usepackage{amssymb}
\usepackage{graphicx}
\usepackage{amsbsy}
\usepackage{subfigure}
\usepackage{color}
\usepackage[toc,page]{appendix}
\usepackage{makeidx}

\makeindex

\journal{ Modelling and Simulation in Materials Science and Engineering}

\begin{document}

\begin{frontmatter}

\title{On the choice of homogenization method   to achieve effective mechanical properties of composites reinforced by ellipsoidal and spherical particles}

\author[1]{Viwanou Hounkpati}
\author[2]{Vladimir Salnikov}
\author[1]{Alexandre Vivet} 
\author[3]{Philippe~Karamian-Surville}

\address[1]{Normandie Univ, ENSICAEN, UNICAEN, CEA, CNRS, CIMAP, 14000 Caen, France}
\address[2]{University of Luxembourg, Mathematics Research Unit
6, rue Richard Coudenhove-Kalergi, L-1359 Luxembourg}

%\address[3]{Normandie Univ, UNICAEN, CNRS, LMNO, 14000 Caen, France;}
\address[3]{Normandie Univ, UNICAEN, CNRS, LMNO, 14000 Caen, France\\
 {viwanou.hounkpati@unicen.fr, vladimir.salnikov@uni.lu, 
  alexandre.vivet@unicaen.fr, philippe.karamian@unicaen.fr}} 

\begin{abstract}

In this paper, several rigorous numerical simulations were conducted to examine  the relevance of mean-field micromechanical models compared to the Fast Fourier Transform  full-field computation by considering spherical or ellipsoidal inclusions. To be more general, the numerical study was extended to  a mixture of  different kind of microstructures consisting of spheroidal shapes within the same RVE.  Although the Fast Fourier Transform  full field calculation is sensitive to high contrasts, calculation time, for a combination of complex microstructures, remains reasonable compared with  those obtained with mean-field micromechanical models. Moreover, 
for low volume fractions of inclusions, the results of the mean-field approximations and those of the Fast Fourier Transform-based (FFTb) full-field computation are very close, whatever the inclusions morphology is. For RVEs consisting of ellipsoidal or a mixture of ellipsoidal and spherical inclusions, when the inclusions volume fraction becomes higher, one observes that Lielens' model and the FFTb full-field computation give similar estimates. The accuracy of the computational methods depends on the shape of the inclusions' and their volume fraction.

\end{abstract}


\end{frontmatter}

%% \linenumbers

\section{Introduction and motivation}
\label{sec:intro}
During the last decades, the micromechanics of heterogeneous materials, which is an area of very fertile research at the boundary between physics and mechanics of materials, evolved from the so-called mean-field approaches~\citep{Zaoui2002, Bonfoh2012} to full-field schemes such as Finite Element Method (FEM)~\citep{Pan2008, Chawla2004, 
Smit1998}  or more recently Fast Fourier Transforms (FFT) methods initiated by Moulinec and Suquet ~\citep{Moulinec1998, Moulinec2003,Moulinec1994}. The aim of these approaches is to predict the local and the effective behaviors of materials which are strongly influenced by their microstructure (constituents' properties, volume fractions, shapes, orientations, etc.). In reinforced composite materials, inclusions can have several morphologies with different geometrical orientations. To accurately describe the effective behavior of such materials, a highly detailed description of the microstructures is required.
The full-field approach used in this paper is the Fast Fourier Transform-based (FFTb) homogenization technique~\citep{Michel2001, Salnikov2015}  where an iterative scheme is used for computing effective properties of each given RVE. One of the main advantages of this approach is its low time and memory consumption in comparison for example with finite element method. The application of the FFTb methods to the study of the overall behavior of materials involves a preliminary step of 3D RVE generation. Another important feature of this method is that the generated RVE well approximates, at least in certain aspects, the real microstructure of the material.
Mean-field homogenization methods rely on a statistical analysis of the microstructure. Several mean-field homogenization models have been developed for predicting the mechanical properties of ellipsoidal fibers reinforced composites (as well as spherical particles reinforced composites), such as the dilute solution of Eshelby~\citep{Eshelby1957}, the Hashin– Shtrikman bounds~\citep{Hashin1963}, the self-consistent scheme~\citep{Hill1965}, the Mori-Tanaka model~\citep{Mori1973,Benveniste1987} and Lielens’'model~\citep{Lielens1998}, among others.
Several works were conducted to highlight the validity domains of these mean-field micromechanical models (by comparing them to full-field approaches) when predicting the elastic properties of various RVEs with different microstructures. Most of these studies have been focused on  spherical particles reinforced composites ~\citep{Ghossein2012,Pierard2004, Marur2004, Kari2007, Klusemann2010, Cojocaru2010}, aligned fibers reinforced composites ~\citep{Tucker1999}, randomly oriented fiber reinforced composites~\citep{Ghossein2014, Mortazavi2013, Kari2007b, Berger2007, Bohm2002} or microstructures with aligned or randomly oriented clay platelets~\citep{Pahlavanpour2013,Sheng2004, Hbaieb2007,Figiel2009}.

This work deals with spherical particles and ellipsodal fibers reinforced composites and is extended to microstructures consisting of both shapes within the same RVE. In the mean-field approaches, these different shapes (sphere and ellipsoid) were taken into account via the Eshelby's tensor. Eshelby \citep{Eshelby1957, Eshelby1961} has shown that the strain field within a homogeneous ellipsoidal inclusion in an infinite elastic matrix is uniform, if the eigenstrain in the inclusion is uniform. He also stated that among finite inclusions the ellipsoidal alone has this convenient property. The absence of the Eshelby's property for non-ellipsoidal property was confirmed by Lubada and Markenscoff \citep{ Lubada1998}. Many works \citep{Zou2010,Klusemann2010, Ru2000, Ru2003} have been conducted considering complex shapes (non-ellipsoidal) using different Eshelby's tensors but these latter were limited to 2D problems. Thus, only spherical and ellipsoidal inclusions (and a mixture) were investigated in this paper, as we are interested in 3D volumes.

 The effective material properties obtained using the numerical FFTb homogenization techniques were compared with three different analytical methods: the normalized self-consistent scheme (NSC)~\citep{Li1999, Hounkpati2014}, the Mori-Tanaka model (MT)~\citep{Mori1973} and Lielens' model~\citep{Lielens1998}. The influence of the inclusions morphology on the accuracy of these homogenization techniques to predict the mechanical properties of reinforced composites was investigated. Since the FFTb method requires a complete description of the RVEs, the numerical results obtained with this method have been 
considered as baseline data   to quantify the discrepancies with analytical methods. The paper is organized as follows: the next section briefly recalls the RVE generation methods proposed in~\citep{Salnikov2015,Salnikov2015b, SVDP}. Then, we present the FFT-based homogenization technique and the mean-field homogenization models used in this work. Afterwards, effective properties predicted by analytical models are  compared to those obtained numerically in order to rigorously validate the investigated models and to highlight the influence of the inclusion volume fraction on their accuracy and their sensibility to the inclusions geometrical orientation.

%\section{Sample generation and computational techniques} 

\section{RVE generation} \label{sec:RVE}

Representative volume elements  are a powerful tool to model media with inhomogeneities. It consists of a representative pattern the size of which 
must respect several criteria. It can be defined as a volume $V$ large enough to provide accurate informations on the microstructure  of the medium~\citep{Hashin1983,Hill1963} but also not too big to remain elementary and limit the computation cost  and respect a minimum scale ratio with the macroscopic material~\citep{Kanit2003}.

 Generating digital samples for computation is a key step in the process of modelling of the behavior of composite materials.  Taking into account the morphology of composites is rather a complex task and this is a nice challenge to design an algorithm to model the network and shape of inclusions. On the one hand, the algorithm should  be able to approach rather complex geometries; on the other hand the algorithm has to be reliable and fast  enough. Roughly speaking the generation step must be shorter than the computation of the effective mechanical properties in itself.

 Many studies were conducted with  inclusions  represented by simple geometric objects like spheres or ellipsoids (see for example,~\citep{segu, levesque_sp, zhao, man}). For more complicated geometry, which is a challenging  task, the problem  is how to  manage the intersection between  inclusions.

 % Essentially two important methods can achieve the task: random sequential adsorption (RSA) and a method based on molecular dynamics (MD). The RSA~\citep{rsa1} is based on sequential addition of inclusions verifying for each of them  the intersection; the main idea of the MD~\citep{LSA, williams, levesque_ell} is to make the inclusions move, until they reach the desired configuration.

 Several approaches which are less or more efficient for various geometries  can be adopted to generate an RVE. Historically the first one is known as RSA which stands for random sequental adsorption (RSA, see for example \citep{zhao, rsa1}) - a series of parameters of the geometry are randomly generated and some verifications are made on these parameters to satisfy some imposed constraints like
 interections. Practically, the algorithm starts with an empty box in which all the inclusions are added one after the other while rejecting those that do not satisfy the constraint of non-intersection. In the case of the RSA algorithm, the process is usually achieved at low volume fraction of inclusions, otherwise the generating process may take some time or even get stuck while the RVE is still far from the desired fraction voume inclusions.

The RSA method needs an efficient algorithm in order to verify  intersection between the geometric objects whereas the second method  based on molecular dynamics  necessitates an algorithm to predict the time of  intersection of moving objects, which exists 
for a very limited class of shapes and often amounts to a difficult minimization problem. More description for  interested reader on  the classical RSA and a time-driven version of MD applied to the  mixture of inclusions of spherical and cylindrical shapes can be found in ~\citep{VDP}.
The key point  in both approaches  is the explicit formulation of algebraic conditions of intersection of a cylinder with a sphere and of two cylinders. (cf. Algorithm 1 in ~\citep{VDP})

\begin{figure}[!ht]  
\centering
 \includegraphics[width=0.47\linewidth]{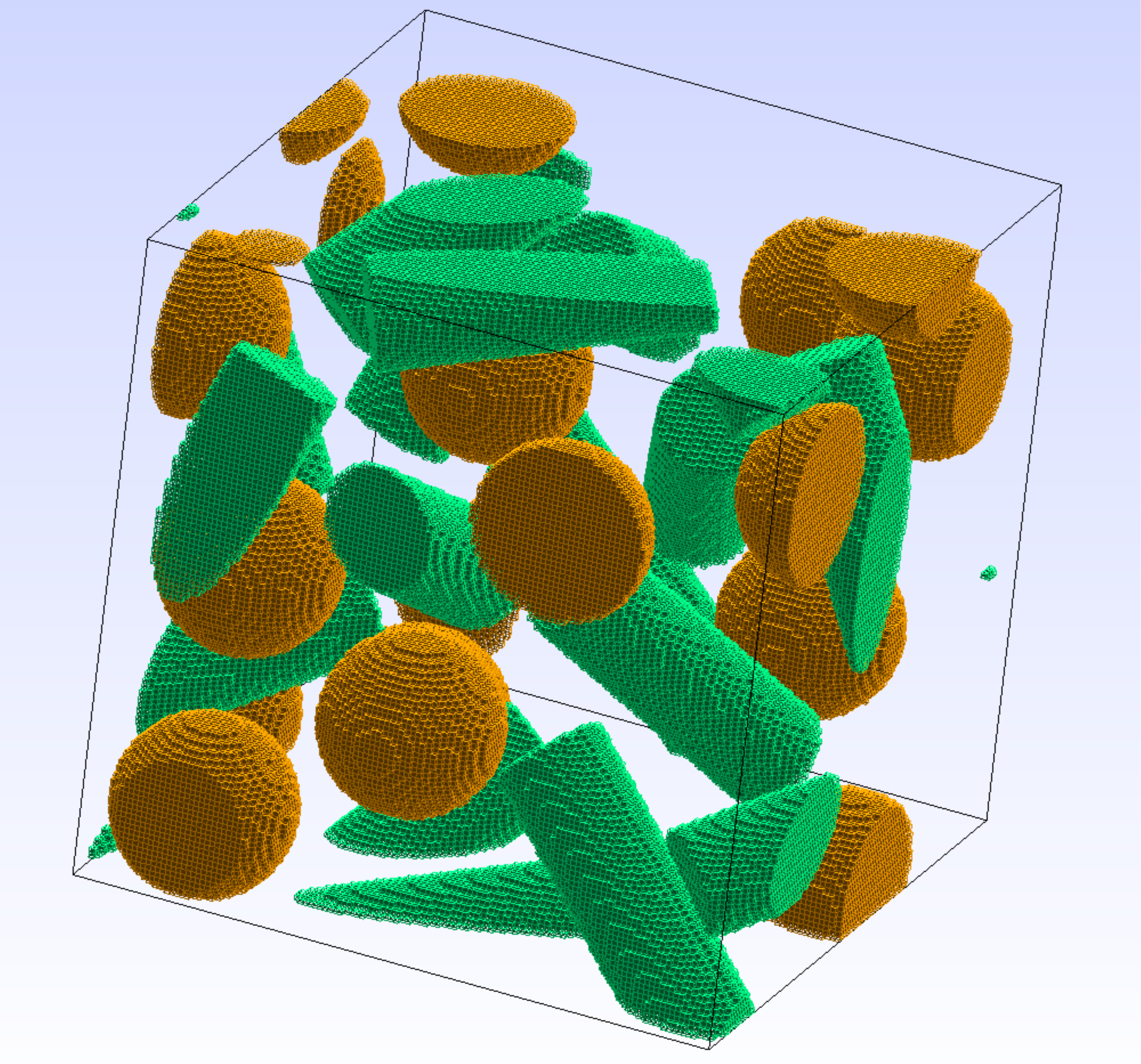}  \includegraphics[width=0.52\linewidth]{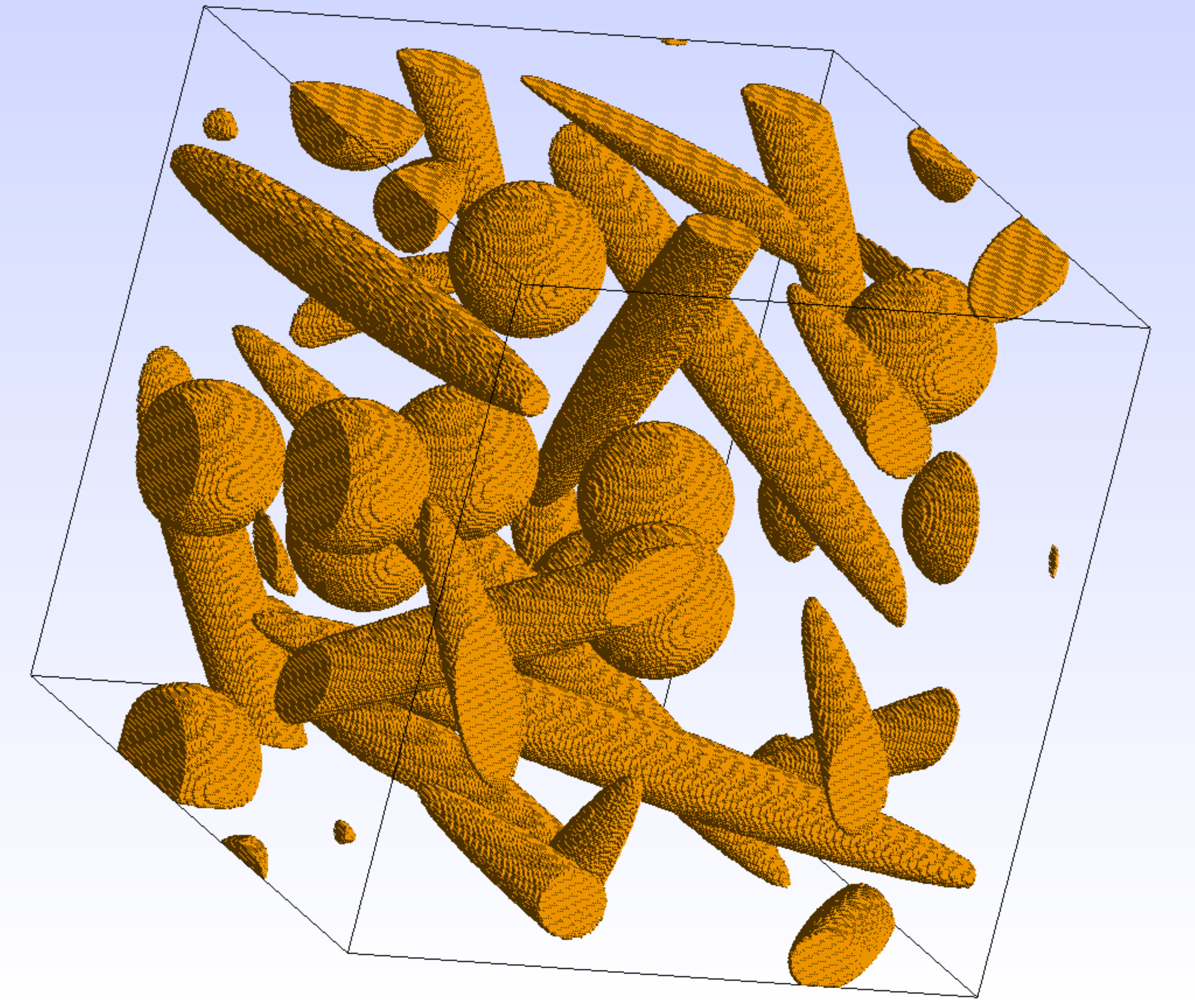} 
\caption{\label{fig:rve3D} 3D view of a generated RVE: spherical-cylindrical and spherical-ellipsoidal inclusions with periodic boundary  conditions.} 
 \end{figure}

The RSA approach is extremely efficient for relatively small volume fractions of inclusions (up to $30 \%$). The MD-based method is powerful for higher volume fractions (of order $50 -60 \%$): MD generates a configuration in about a second whereas the RSA can get stuck. 
Figure \ref{fig:rve3D} exhibits two samples with a mixture  of non-intersecting spherical-cylindrical
and spherical-ellipsoidal inclusions.

The outcome of these algorithms is a list of data of all  inclusions such as  coordinates of centers, radii, and eventually axes of symmetry of inclusions.  This is appropriate for FFT-based homogenization procedures applied to the pixelized samples, as well as finite element 
computations on the mesh constructed from this pixelization.

In the case of molecular dynamics (MD) another approach is used: Basically, in the first step, all inclusions are added simultaneously and they are allowed to interact with each other, which means that the principle of non-intersecting is violated. The algorithm starts moving the inclusions such that they get their right place in the box  so that all the inclusions satisfy the non-intersecting contraint. In \citep{Salnikov2015b} we explain how the above mentioned methods can be used in order to generate the RVEs with cylindrical and spherical inclusions. 
We describe in details the random generation of non-intersecting inclusions, as well as the relaxation procedure allowing us to produce non-intersecting configurations from the intersecting ones. The section 2 in \citep{Salnikov2015b}  focuses on the intersection conditions of two kinds of geometries : Sphere with cylinder and the case of two cylinders.  Moreover the format of the output of the algorithms is very convenient, namely all the information  about the RVE are  encoded in the concise vector form. We can on the one hand pixelize it to 
have a natural discretization of the analyzed sample, and on the other hand keep track of 
orientations of the inclusions.

 In the sequel, in the prospect of a large number of RVEs we have chosen to generate them with the help of  RSA or MD based procedures.  The inclusion network is generated according to different morphological 
features such as the orientation, the aspect ratio or the dispersion of ellipsoids.

\section{FFT-based homogenization scheme} \label{sec:FFT}
The previous paragraph has  briefly presented the two main procedures to generate RVEs. Considering  an efficient scheme to generate samples of RVES we can now proceed to the evaluation of the effective mechanical properties of reinforced composites.  In  practice, we set some parameters of the material such as volume fraction, shape of inclusion like spheres, ellispoids and the aspect ratio of inclusion. We then generate a database of samples of RVEs of composite material with these parameters. One should keep in mind that all these parameters can be  randomly  managed. Thereafter one can perform an accurate computation of the effective properties on this database of RVEs with the help of a deterministic homogenization technic either using a classical finite element method analysis for example or solving the Lippmann-Schinger equation with the FFT technique for instance. In the sequel and in the present study, we focus on particular RVEs  for each different situation examined and for which the RVEs are supposed to be a typical sample that can be encountered in practical situations.

Let us now describe the homogenization procedure.  Let $V$ be  a volume element, and let us note $\mathbf{u}(\mathbf{x})$  the displacement  field defined at any point $\mathbf{x} \in V$. Let  $\varepsilon(\mathbf{x}) = \varepsilon(\mathbf{u}(\mathbf{x})) = \frac{1}{2}(\nabla\mathbf{u}(\mathbf{x}) + \nabla\mathbf{u}(\mathbf{x})^T)$ be  the strain tensor in the model of small deformations, and let $\sigma(\mathbf{x})$ be the stress tensor, subject to the condition $\mathrm{div}\, \sigma(\mathbf{x}) = 0$. In the linear case we have
$\sigma(\mathbf{x}) = c(\mathbf{x})\!:\!\varepsilon(\mathbf{x})$ with the stiffness tensor $c(\mathbf{x})$.\\

Note that for a composite material, the stiffness tensor depends on the point $\mathbf{x}$: 
which belongs to the matrix or to the inclusion.
We assume that the averaged strain $<\!\varepsilon\!> = E$ is prescribed, and decompose $\varepsilon(\mathbf{x})$ in two parts: $\varepsilon(\mathbf{u}(\mathbf{x})) = E + \varepsilon(\mathbf{\tilde u}(\mathbf{x}))$, which is 
equivalent to representing  $\mathbf{u}(\mathbf{x}) = E.\mathbf{x} + \mathbf{\tilde u}(\mathbf{x})$, 
for $\mathbf{\tilde u}(\mathbf{x})$ being periodic on the boundary of $V$.  
Hence, the mechanical problem which has to be solved  reads
\begin{eqnarray} \label{pb1}
  \sigma (\mathbf{x}) = c(\mathbf{x}) : (E + \varepsilon(\mathbf{\tilde u}(\mathbf{x})), 
  \quad \mathrm{div}  \, \sigma(\mathbf{x}) = 0, \\
  \mathbf{\tilde u}(\mathbf{x}) \,\, \mathrm{ periodic}, \, 
  {\sigma}(\mathbf{x}).\mathbf{n} \,\,\mathrm{ antiperiodic}.\nonumber 
\end{eqnarray}
The solution of (\ref{pb1}) is the tensor field $\sigma(\mathbf{x})$, taking its average 
one can obtain the homogenized stiffness tensor $c_{hom}$ from the equation 
\begin{equation}\label{average}
<\!\sigma(\mathbf{x})\!> = c_{hom} : <\!\varepsilon(\mathbf{x})\!>
\end{equation}

To obtain all the components 
of $c_{hom}$ in 3-dimensional space one needs to perform the computation of $<\!\sigma(\mathbf{x})\!>$ for six 
independent deformations $E$, which morally correspond to usual stretch and shear tests. 
In order to solve the problem (\ref{pb1}): one can employ the finite elements method, or discretize (basically voxelize) the RVE to use  the FFT-based homogenization scheme~\citep{moul-suq, monchiet}. Our choice has gone to the last one. In the case of a homogeneous isotropic material,  the equations of (\ref{pb1}) have in the Fourier space a  Green operator  and basically produce an exact solution. The computation leads to an iterative procedure described and summerised  in ~\citep{MMS, eyre, Salnikov2015}.Once the above algorithm has converged, we compute $<\!\sigma(\mathbf{x})\!>$ which has to be inserted into the equation (\ref{average}).

\section{Mean-field homogenization models}
In this section, we briefly describe the mean-field homogenization models investigated in this work. A detailed description can be found in the references mentioned in the texts.

\subsection{Preliminaries: accounting for the inclusions' geometrical orientation}

The main idea of homogenization models is to find a globally homogeneous medium equivalent to the original composite. We restrict the composite to the matrix-inclusion type with perfect interfacial bonds between inclusions and their immediate surrounding matrix. Because the  local frame of reference $Rl$ attached to  the inclusions' main axes are not always identical to the global one $Rg$ attached to RVE coordinate system, subscripts $Rl$ and “$Rg$” are added, for the sake of clarity, to each tensor to indicate the coordinate system in which it is expressed. One can note that the subscripts "$Rl$" and “$Rg$” are not necessary when the inclusions are spherical or all aligned, because in this case $Rl$ and $Rg$ are the same. For randomly oriented inclusions reinforced composites, the geometrical orientation of each inclusion is described by three Euler angles
$\phi_1,\phi,\phi_2$. The transition between the local coordinate system $Rl$ of the inclusion and the RVE
system $Rg$ is made by these three Euler angles in the Bunge convention~\citep{Bunge1982} where the transformation matrix is given by $m(\phi_1,\phi,\phi_2)$ defined as follows.
\begin{center}

\begin{displaymath}
%m(\phi_1,\phi,\phi_2)=
\begin{bmatrix}
  \cos\phi_1 \cos\phi_2 - \sin\phi_1 \sin\phi_2\cos\phi&  -\cos\phi_1 \sin\phi_2 - \sin\phi_1 \cos\phi_2\cos\phi & \sin\phi_1\sin\phi\\ 
   \sin\phi_1 \cos\phi_2 + \cos\phi_1 \sin\phi_2\cos\phi&-\sin\phi_1 \sin\phi_2 + \cos\phi_1 \cos\phi_2\cos\phi  &- \cos\phi_1\sin\phi\\
  \sin\phi_2\sin\phi & \cos\phi_2\sin\phi&\cos\phi
\end{bmatrix}
\end{displaymath}
\end{center}

This matrix is applied, for a given rank four tensor K, as follows:

\begin{eqnarray*}
_{Rg}K_{ijkl}=m_{im}m_{jn}m_{ko}m_{lp}  (_{Rl}K_{mnop})
\end{eqnarray*}

The homogenized stiffness tensor $C$ of a composite consisting of n different types of inclusions (in terms of shape and geometrical orientation) in the RVE coordinate system Rg is given by

\begin{eqnarray*}
C=<~\!_{Rg}C : ~\!_{Rg}A>_{RVE}=f_m ~\!_{Rg}C_m : ~\!_{Rg}A_m +\sum_{i=1}^{n}f_i ~\!_{Rg}C_i:~\! _{Rg}A_i,
\end{eqnarray*}

\noindent where $f_i$, $C_i$ and $A_i$ denote respectively the volume fraction, the stiffness tensor and the strain-localization tensor of inclusions exhibiting the same shape and the same geometrical orientation. $f_m$, $C_m$ and $A_m$ denote respectively the volume fraction, the stiffness tensor and the strain- localization tensor of the matrix. $A:B$ denotes the double scalar product using the
Einstein summation convention. $<\bullet>_{RVE}$ stands for the volume average over the whole RVE (matrix + inclusions). From the average strain theorem~\citep{Hill1963}, it can be verified that:

\begin{eqnarray*}
<  ~\!_{Rg}A>_{RVE}=f_m  ~\! _{Rg}A_m +\sum_{i=1}^{n}f_i  ~\!_{Rg}A_i = I
\end{eqnarray*}

\noindent  where $I$ is the fourth-order unit tensor. By replacing in relation (3) $f_m ~\!_{Rg} A_m$ by its expression deduced from (4), one can obtain:

\begin{eqnarray*}
C=_{Rg}C_m+\sum_{i=1}^{n}f_i ( _{Rg}C_i - _{Rg}C) : _{Rg}A_i 
\end{eqnarray*}

This is the main relation used to calculate the homogenized stiffness tensor. The strain-localization tensor $A_i$ in this relation differs from one model to another. The next sections present the expression of the strain- localization tensor of the mean-field models studied in this paper.

 \subsection{Normalized self-consistent model}
 
 The self-consistent model (SC) assumes that each inclusion is embedded in a fictitious homogeneous matrix possessing the composite's unknown stiffness $C$. Generally, the model gives good predictions for polycrystals but is less accurate in the case of certain composites. The strain-localization tensor for the self-consistent model is written $Rl$ as follows~\citep{Hill1965} :
 
\begin{eqnarray*}
_{Rl}A_i^{SC} = [I +_{Rl}E : (_{Rl}C_i - _{Rl}C_m)]^{-1} 
\end{eqnarray*}

 $E$ is the Morris' tensor~\citep{Morris1970}, which represents the interaction between an inclusion with a given morphology and the homogeneous equivalent medium. In the case of an ellipsoidal inclusion whose principal axes lengths are $\{2a_1,2a_2, 2a_3\}$, it is written in the coordinate system of the inclusion $Rl$ as follows:
 
\begin{eqnarray*}
~\!_{Rl}E_{ijkl} =\frac{1}{4\pi}\int_0^\pi\sin\theta d\theta\int_0^{2\pi}\gamma_{ijkl}d\phi = ~\!_{Rl}S_{ijkl}^{Esh}~\!_{Rl}C_{ijkl}^{-1}
\end{eqnarray*}

with

\begin{eqnarray*}
\gamma_{ikjl} &=& K^{-1}(\xi)\xi_j\xi_l,\\
K_{jp}(\xi)&=&C_{ijpl}\xi_i\xi_l,\\
 \xi_1&=&\frac{\sin\theta\cos\phi}{a_1},\\
 \xi_2&=&\frac{\sin\theta\sin\phi}{a_2}, \\
  \xi_3&=&\frac{\cos\theta}{a_3}.
\end{eqnarray*}

\noindent where $S^{Esh}$ is the Eshelby tensor; $a_1, a_2, a_3$ are used to describe the shape of inclusions.
It was shown by Li~\citep{Li1999} that all micromechanical approaches should provide a diagonally symmetric stiffness tensor, give identical thermal and mechanics stress tensors using two equivalent methods, satisfy the internal-consistency relationships between the effective moduli, and exhibit the correct behaviors at dilute and unitary concentration limits. Unfortunately, the application of traditional scale transition models such as self-consistent models or Mori-Tanaka to the case of multi-phase materials containing heterogeneities of different shapes does not simultaneously satisfy all these five fundamental criteria. It often occurs that the first three criteria are not fulfilled~\citep{Li1999, Lacoste2010, Benveniste1991}. To overcome all these several theoretical difficulties, Li~\citep{Li1999} proposed an effective-medium-field micromechanics approximation using normalized strain localization tensor. Under this approximation, the strain localization tensor can be written, in the inclusion's frame of reference $Rl$ attached to its main axes, as~\citep{Li1999,Hounkpati2014}:%~\citep{Li1999,Hounkpati2014, Hounkpati2015}:

\begin{flalign*}
~\!_{Rl}A_i^{NSC} =&~\!_{Rl}A_i^{SC} : <~\!_{Rl}A_i^{SC} >_{RVE}^{-1}\\
 ~\!_{Rl}A_i^{NSC} =&[I +_{Rl}E : (~\!_{Rl}C_i - ~\!_{Rl}C)]^{-1}: <[I +~\!_{Rl}E : (~\!_{Rl}C_i - ~\!_{Rl}C)]^{-1}>_{RVE}^{-1}
\end{flalign*}

This is the strain localization tensor for the so-called normalized self-consistent model (NSC) [34,41] which implementation requires an iterative loop in order to determine the homogenized stiffness tensor of the composite, using relation (5).

 \subsection{Mori-Tanaka model}
 
 In the Mori–Tanaka model ~\citep{Mori1973}, the strain localization tensor is given by the expression:
 
 \begin{eqnarray*}
~\!_{Rl}A_i^{MT} =~\!_{Rl}A_i^{dilute} : < ~\!_{Rl}A_i^{dilute} >_{RVE} ^{-1}
\end{eqnarray*}

\noindent where $A^{dilute}$ is the dilute strain localization tensor in the case of a single particle in an infinite i
matrix, given by:
 
\begin{eqnarray*}
 ~\!_{Rl}A_i^{dilute} =[I +~\!_{Rl}E : (~\!_{Rl}C_i -~\! _{Rl}C_m)]^{-1}
\end{eqnarray*} 
 
 In this relation, a special attention should be paid to the Morris' tensor $E_m$ which is not, here,
function of the homogenized stiffness tensor $C$, but function of the matrix stiffness tensor $C_m$.
It means that the Morris' tensor $E_m$ has the same expression like in equations (7) and (8) with
$C$ replaced by $C_m$. In the other words, the Morris' tensor is computed by using the matrix
properties as the infinite media. The strain localization tensor $~\!_{Rl}A_i^{dilute}$ is equal to the unit
tensor for the matrix (when i = m). By replacing $A_i^{MT}$ in equation (5), one can calculate the homogenized stiffness tensor of the composite. By applying carefully the change between $Rl$ and $Rg$, the Mori-Tanaka model delivers symmetric effective stiffness tensors whatever the microstructure of the composites investigated in this work.

 \subsection{Lielens' model}
 Lielens and his co-workers~\citep{Lielens1998} proposed an homogenization model which is a nonlinear interpolation between the Mori-Tanaka and inverse Mori-Tanaka method and between the Hashin-Shtrikman bounds, respectively, for aligned reinforcements. More precisely, this model interpolates the inverse of the strain-localization tensor between the case where the stiffest phase is embedded in the more compliant phase and that where the most compliant phase is embedded in the stiffest phase ~\citep{Tucker1999}. The strain localization tensor is given by:
 
 \begin{eqnarray*}
~\!_{Rl}A_i^{Lielens} = ~\!_{Rl}A_i^{*}: <~\!_{Rl}A_i^{*}>^{-1}_{RVE}
\end{eqnarray*} 

where

\begin{eqnarray*}
~\!_{Rl}A_i^{*} = [(1-f)(~\!_{Rl}A_i^{lower})^{-1} +f (~\!_{Rl}A_i^{upper})^{-1}]^{-1}
\end{eqnarray*} 

\noindent f is the interpolating factor which depends on the inclusions volume fraction ~\citep{Lielens1998,Tucker1999}. For a set of inclusions with the same shape (ellipsoid, sphere,...) having a volume fraction $f_{shape}$, the interpolating factor is given by:

\begin{eqnarray*}
f= \frac{1}{2}f_{shape}(1+f_{shape})
\end{eqnarray*} 

In our study, for the RVEs exhibiting inclusions with two different shapes (ellipsoid and
sphere), an interpolating factor is considered for each shape according to their volume
fraction. $A_i^{lower}$ and $A_i^{upper}$ are given by:

\begin{eqnarray*}
~\!_{Rl}A_i^{lower}= [I +~\!_{Rl}E_m : (_{Rl}C_i - ~\!_{Rl}C_m)]^{-1} \\ 
~\!_{Rl}A_i^{upper}= [I +~\!_{Rl}E_i : (_{Rl}C_m - ~\!_{Rl}C_i)]^{-1}
\end{eqnarray*} 

Note that in these relations, the Morris' tensors $E_m$ and $E_i$ are computed by using the matrix
and the inclusion properties respectively as the infinite media.

\section{ Computation of composite effective mechanical properties: results and discussion}

The main purpose of this study is to highlight the influence of the inclusions' morphology on the accuracy of the prediction of reinforced composites mechanical behaviour. Several RVEs exhibiting different morphologies  or geometrical orientations have been studied. The accuracy of homogenization models was evaluated for composites made of an isotropic matrix reinforced with isotropic spherical and/or ellipsoidal particles. Effective properties predicted by mean-field homogenization analytical models have then been compared to those obtained numerically (by FFT-based homogenization technique) in order to  validate the investigated models and to highlight the influence of the inclusions' volume fraction on their accuracy. Note that the mean-field approaches investigated in this work (normalized self-consistent, Mori-Tanaka and Lielens' models) provide a diagonally symmetric stiffness tensor and satisfy the internal-consistency relationships between the effective moduli, regardless of the RVE.
The homogenized stiffness tensor C of each RVE studied was not strictly isotropic because a finite number of inclusions is considered and/or these inclusions exhibit different shapes or geometrical orientations. The effective bulk $K_{eff}$,  shear $\mu_{eff}$  and Young's moduli $E_{1eff}$  (in direction 1 of the sample) were computed for each RVE using the corresponding homogenized stiffness tensor (coefficient $C_{ijkl}$) or matrix (coefficient $C_{IJ}$):

\begin{flalign*}
K_{eff}&=\frac{C_{iijj}}{9},\\
\mu_{eff}&=\frac{3C_{ijij}-C_{iijj}}{30},\\
 E_{1eff} &=\frac{1}{S_{11}}=\frac{C_{11}C_{22}C_{33}+2C_{12}C_{23}C_{31}-C_{12}^2C_{33}-C_{23}^2C_{11}-C_{13}^2C_{22}}{C_{22}C_{33}-C_{23}^2}.
\end{flalign*}

\noindent where S is the inverse of the homogenized stiffness matrix. We were interested in the normalized values of these effective moduli: $\displaystyle\frac{K_{eff}}{K_{m}} $, $\displaystyle\frac{\mu_{eff}}{\mu_{m}} $ and $\displaystyle\frac{E_{1eff}}{E_{m}} $, respectively. The Poisson ratio is assumed   to be $\nu = \frac{1}{3}$ for all phases (inclusions and matrix). 
It was mentioned earlier in this paper that the RSA and MD-based generation algorithms give as out-come a list of inclusions in the “vector” form, i.e. a list of coordinates of centers, radii, and eventually axes of symmetry of the inclusions. From these values, we determined the input parameters of the models:
\begin{enumerate}

\item The radii enable us to determine the aspect ratio $\displaystyle \frac{a_1}{a_2}$ of the ellipsoidal inclusions defined by theirs principal axes lengths  $a_1, a_2, a_3$ where $a_1>a_2$ and $a_2 = a_3$.
\item From the coordinates of the axes of symmetry of the ellipsoidal inclusions, we determined the three Euler angles $(\phi_1,\phi,\phi_2)$  enabling the transformation between the inclusions frame of reference $Rg$ attached to their main axes and the RVE coordinate system $Rl$.
Note that mean-field approaches do not take into account the position of the inclusion in the RVE. Consequently, the coordinates of inclusions' centers were not exploited by these approaches.
\end{enumerate}

\subsection{Models validity: influence of the inclusions morphology}

We performed here studies to determine the influence of the inclusions' morphology properties on the effective material properties. We studied the influence of the inclusions' morphology on the models' accuracy or validity. Therefore, different RVEs in terms of inclusions' shape were investigated. Table~\ref{table1}    gives a description of the RVEs for which results are presented in this section.

Using these RVEs, we computed the corresponding effective properties by the means of FFT-based homogenization scheme and the mean-field approaches.To quantify the discrepancy between the mean-field models and the FFTb technique  when predicting the effective behaviour of these RVEs, we calculated the relative error.

For each mean-field model, the relative deviations were computed for the normalized effective moduli $\displaystyle\frac{K_{eff}}{K_{m}} $, $\displaystyle\frac{\mu_{eff}}{\mu_{m}} $ and $\displaystyle\frac{E_{1eff}}{E_{m}} $, respectively as follows:

\begin{flalign*}
\displaystyle \delta K &= 100\times\frac{\displaystyle(\frac{K_{eff}}{K_{m}})_{mean field} -\displaystyle(\frac{K_{eff}}{K_{m}})_{FFTb}}{\displaystyle(\frac{K_{eff}}{K_{m}})_{FFTb}},\\
\displaystyle \delta \mu &= 100\times\frac{\displaystyle(\frac{\mu_{eff}}{\mu_{m}})_{mean field} -\displaystyle(\frac{\mu_{eff}}{\mu_{m}})_{FFTb}}{\displaystyle(\frac{\mu_{eff}}{\mu_{m}})_{FFTb}}, \\
\displaystyle \delta E & = 100\times\frac{\displaystyle(\frac{E_{1eff}}{E_{m}})_{mean field} -\displaystyle(\frac{E_{1eff}}{E_{m}})_{FFTb}}{\displaystyle(\frac{E_{1eff}}{E_{m}})_{FFTb}}.
\end{flalign*}

Positive relative deviations reflect an overestimation of the concerned effective properties and negative relative deviations an underestimation.

\begin{table}[!h]
\centering

\small
  \begin{tabular}{|l| l l | l l l|}
    \hline
    \multirow{2}{*}{} &
      \multicolumn{2}{c}{Spheres} &
      \multicolumn{3}{|c|}{Ellipsoidal fibers}\\ 
      \hline
  RVE      & Number & Volume fraction (\%) & Number & Aspect ratio & Volume fraction (\%) \\
    \hline
RVE$_1$& 10 & 5 &10& 10& §6.7\\
RVE$_2$& 0 & 0 &10& 10& §9.8\\
RVE$_3$& 0 & 0 &10& 5& §30\\
RVE$_4$& 2 & 5 &2& 5& §6.7\\
RVE$_5$& 10 & 50 &0& 0& §0\\
\hline
  \end{tabular}
  \caption{Description of the RVEs (unit cells) generated by the RSA and MD-based generation algorithms to highlight the influence of the inclusions' morphology. }\label{table1}
\end{table}
%\footnotetext{stands for Not Available}

For the mean-field computation, a program written under Mathematica software was used for efficiency and fast implementation rather than coding all required procedures in C. The same program coded in C may split the time consumption in 2 or 3 at the cost of  a huge programming time. Besides, the calculation of the effective properties were performed with a 16 GB RAM intel Xeon CPU E5-1650. For the full-field computation a specific program code in C  was used and the calculation were performed with a single processor computational cluster of Intel Nehalem X5560 where only  3 cores and 6 GB Ram were required to compute the effective properties for a given contrast comprised beetwen [10 -100]. All the results  are given in table \ref{table2} for the RVEs listed in \ref{table1}. One can observe that mean-field approaches are more time-consuming than FFTb. This time consumption of the mean-field approaches generically is mainely due to the calculation of the Morris tensor of each geometrical orientation of the ellipsoid fibers. In other words, the computation time increases with the number of ellipsoidal fiber geometrical orientations. 
\begin{table}[!h]
\centering

\small
  \begin{tabular}{l l l l l l l}
    \hline
   & RVE1 & RVE2 & RVE3& RVE4& RVE5& CPU\\
\hline
Mori-Tanaka& 116  & 100& 101&44& 0.03&  intel Xeon\\

Lielens &  134 & 136& 144&33 &0.08& intel Xeon \\

Normalized S-C & 151  &148 &183 &48 &0.83& intel Xeon \\
\hline
FFTb &[ 6--18]   & [ 6--18]& [ 6--22]& [ 6--20]& [ 6--20]& Intel Nehalem\\
\hline
  \end{tabular}
  \caption{CPU times (in minutes) required to compute the effective properties for  contrasts between 10 and 100}\label{table2}
\end{table}

\begin{figure}[!ht]  
\centering
 \includegraphics[width=1\linewidth]{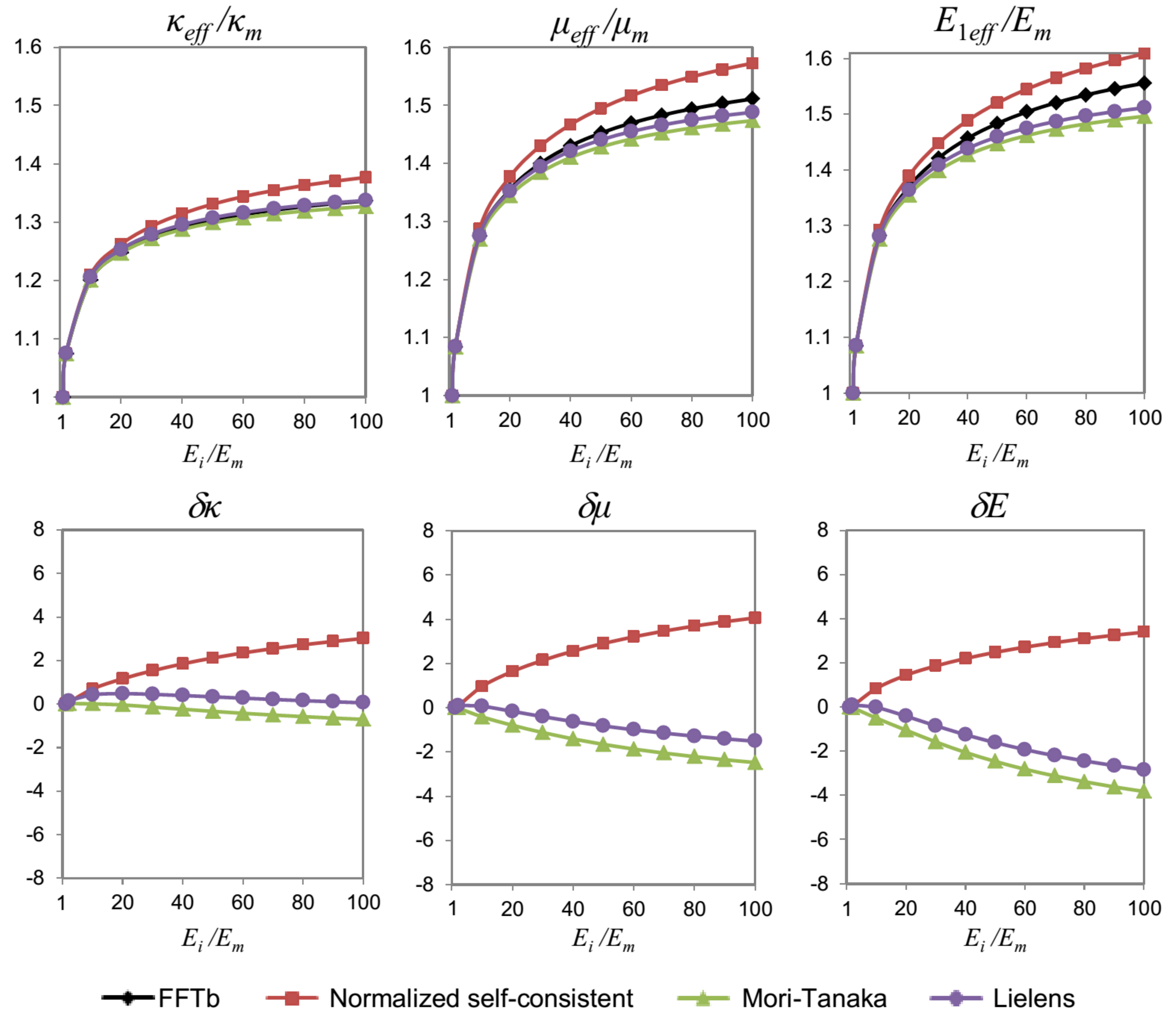}  
 
\caption{\label{fig:results1} Normalized effective bulk, shear and Young's moduli (in direction 1) as function of the Young's modulus contrast for the RVE$_1$ and the corresponding relative deviations (in \%) from the FFTb results.} 
 \end{figure}

Fig. \ref{fig:results1} shows the normalized effective moduli obtained for RVE$_1$ containing a mixture of spherical and ellipsoidal inclusions where the volume fraction of the spherical inclusions is 5\% and 6.7\% 
for the ellipsoidal inclusions. 10 different geometrical orientations were considered for the ellipsoidal inclusions. The aspect ratio of the ellipsoids for this RVE is set at 10. For the considered RVE, the models are in good agreement for all the moduli investigated. However, Lielens' model delivers, for the volume fractions investigated, the lowest gap in approximation results when the RVE contains a mixture of spherical and ellipsoidal inclusions. The Mori-Tanaka's model is also  a good candidate but less than Lielens'.  Although the normalized self-consistent model is the one that gives the biggest difference with the others models, the maximum relative deviation from the FFTb results observed for this latter does not exceed 4.1\%. One can notice that for the normalized Young's  modulus in direction 1, the normalized self-consistent model is slightly more accurate than Mori-Tanaka's for this RVE. For example, at contrast 100, the relative deviations $\delta E$ are -2.8, 3.4 and -3.8\%  for Lielens, Mori-Tanaka and normalized self-consistent models, respectively. 

\begin{figure}[!ht]  
\centering
 \includegraphics[width=1\linewidth]{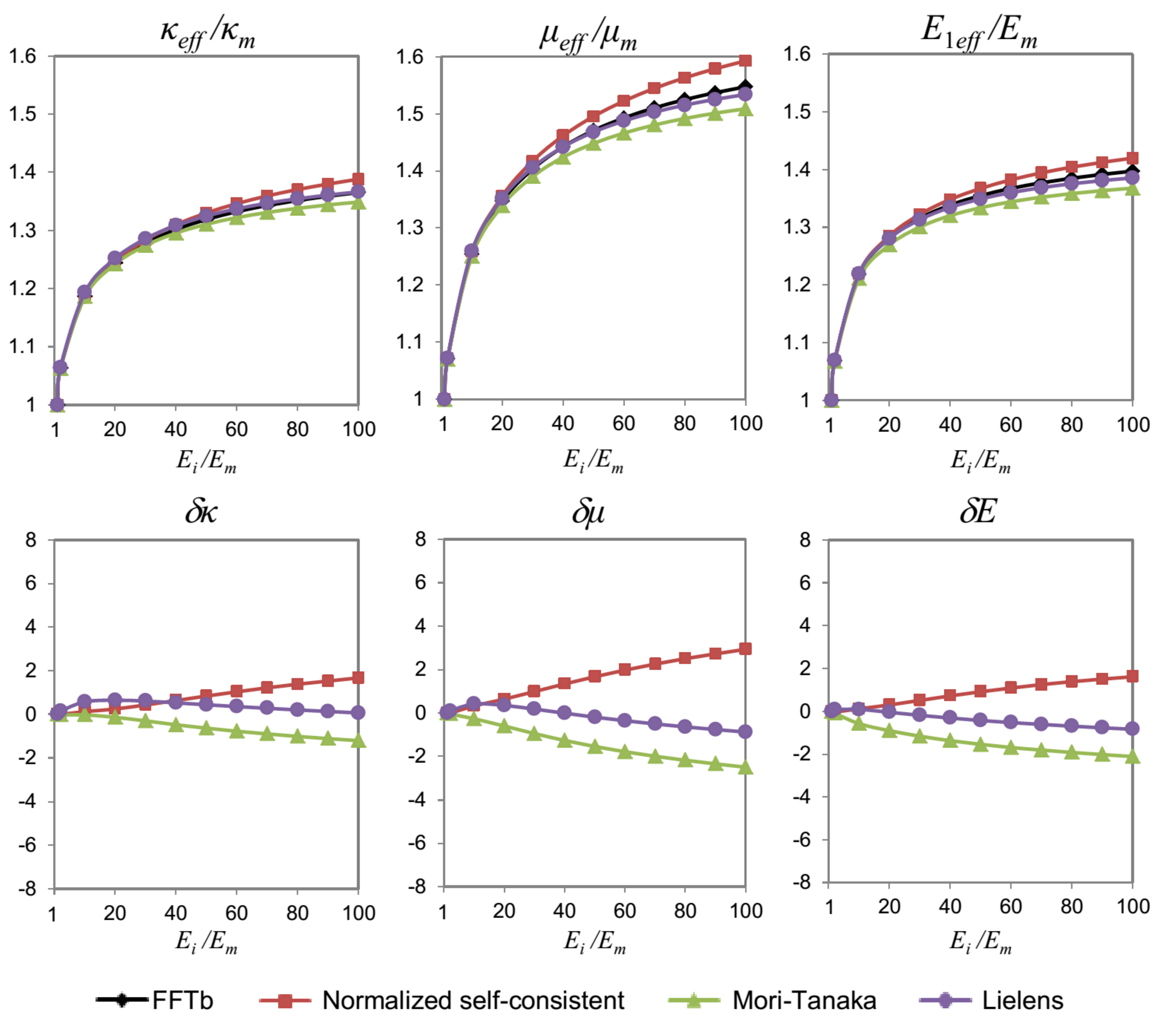}  
 
\caption{\label{fig:results2} Normalized effective bulk, shear and Young's moduli (in direction 1) as function of the Young's modulus contrast for the RVE$_2$ and the corresponding relative deviations (in \%) from the FFTb results.} 
 \end{figure}

The normalized effective shear, bulk and Young's moduli (in direction 1) obtained for RVE$_2$ containing 9.8\% of ellipsoidal inclusions in volume are shown in Fig. \ref{fig:results2}. As RVE$_1$, 10 different geometrical orientations were considered for the ellipsoidal inclusions. Note that the orientation tensors of these RVEs are not the same. The aspect ratio is still set at 10. All the mean-field models provide a good estimate for the shear, bulk moduli and Young's moduli, whatever the Young's modulus contrast of the RVE. Without the spherical inclusions, the accuracy of the models was improved, by comparison with the RVE$_1$. For example, the maximal relative deviation observed on the estimation of the effective Young's modulus at contrast 100 is -3.8\% for RVE$_1$ and only -2.1\% for the RVE$_2$. For this latter, the prediction of the effective bulk and Young's moduli give lesser discrepancy than that of the effective shear modulus.

\begin{figure}[!ht]  
\centering
 \includegraphics[width=1\linewidth]{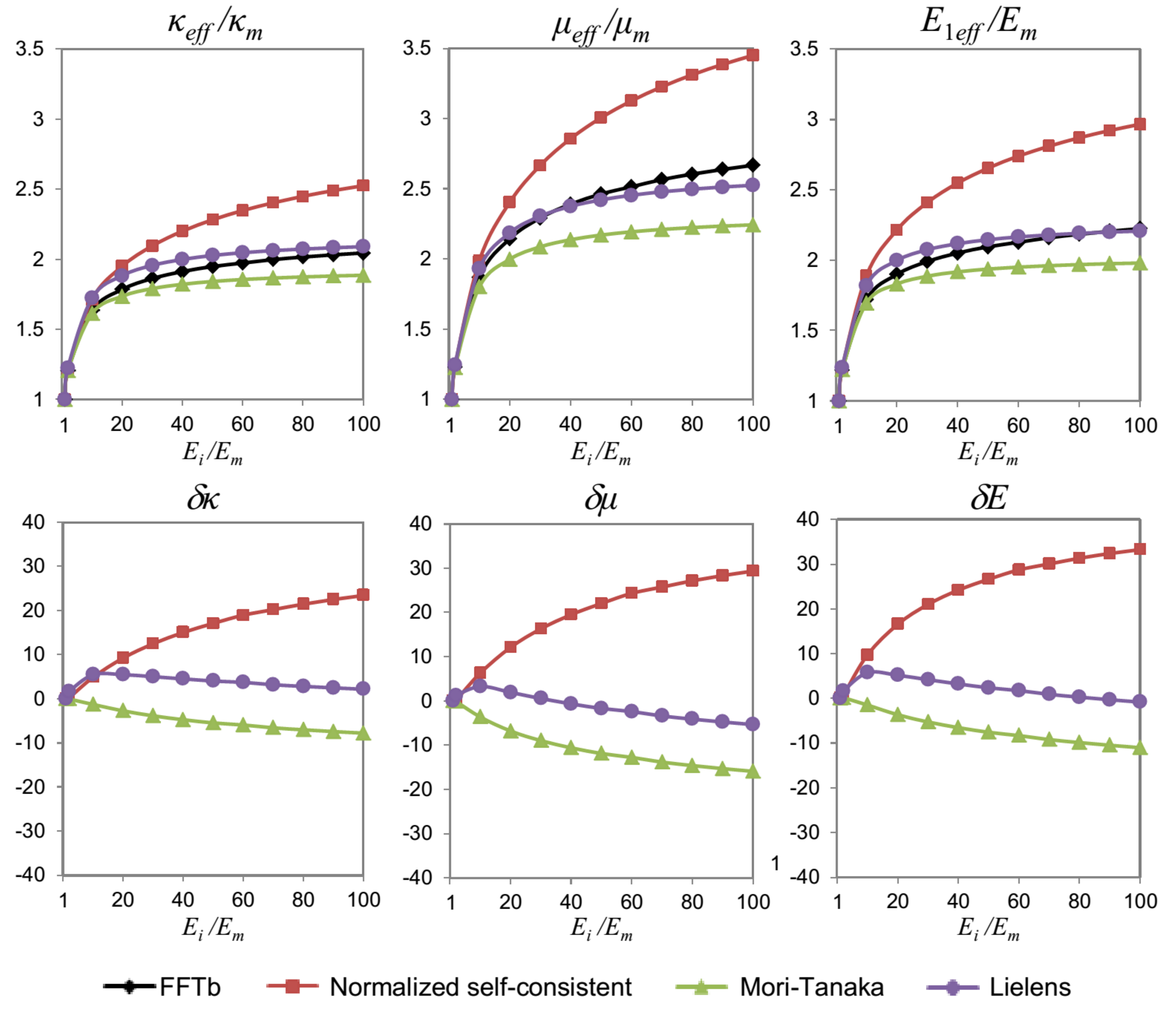}  
 
\caption{\label{fig:results3} Normalized effective bulk, shear and Young's moduli (in direction 1) as function of the Young's modulus contrast for the RVE$_3$ and the corresponding relative deviations (in \%) from the FFTb results.} 
 \end{figure}

We consider now the RVE$_3$ containing 30\% of ellipsoidal inclusions in volume. 10 ellipsoidal inclusions exhibiting different geometrical orientations were generated within the same unit cell as the first two RVEs. In this case, the aspect ratio is set at 5. The normalized effective shear, bulk and Young's moduli (in direction 1) obtained for this RVE are shown in Fig. \ref{fig:results3}. The mean-field models do not provide a good estimate of the effective moduli in this case, except Lielens' model for which all the effective properties were well reproduced whatever the Young's modulus contrast of the RVE. The relative deviations obtained for this model do not exceed 6\%. For Mori-Tanaka model, the maximal relative deviation was observed is -15.9\% on the effective shear modulus and 33.3\% on the effective Young's modulus for the normalized self-consistent model. These significant deviations are presumably due to the high volume of ellipsoidal inclusions. To further investigate the influence of the volume fraction of ellipsoidal inclusions on the accuracy of the mean-field models, a complete study is proposed in the next section.

\begin{figure}[!ht]  
\centering
 \includegraphics[width=1\linewidth]{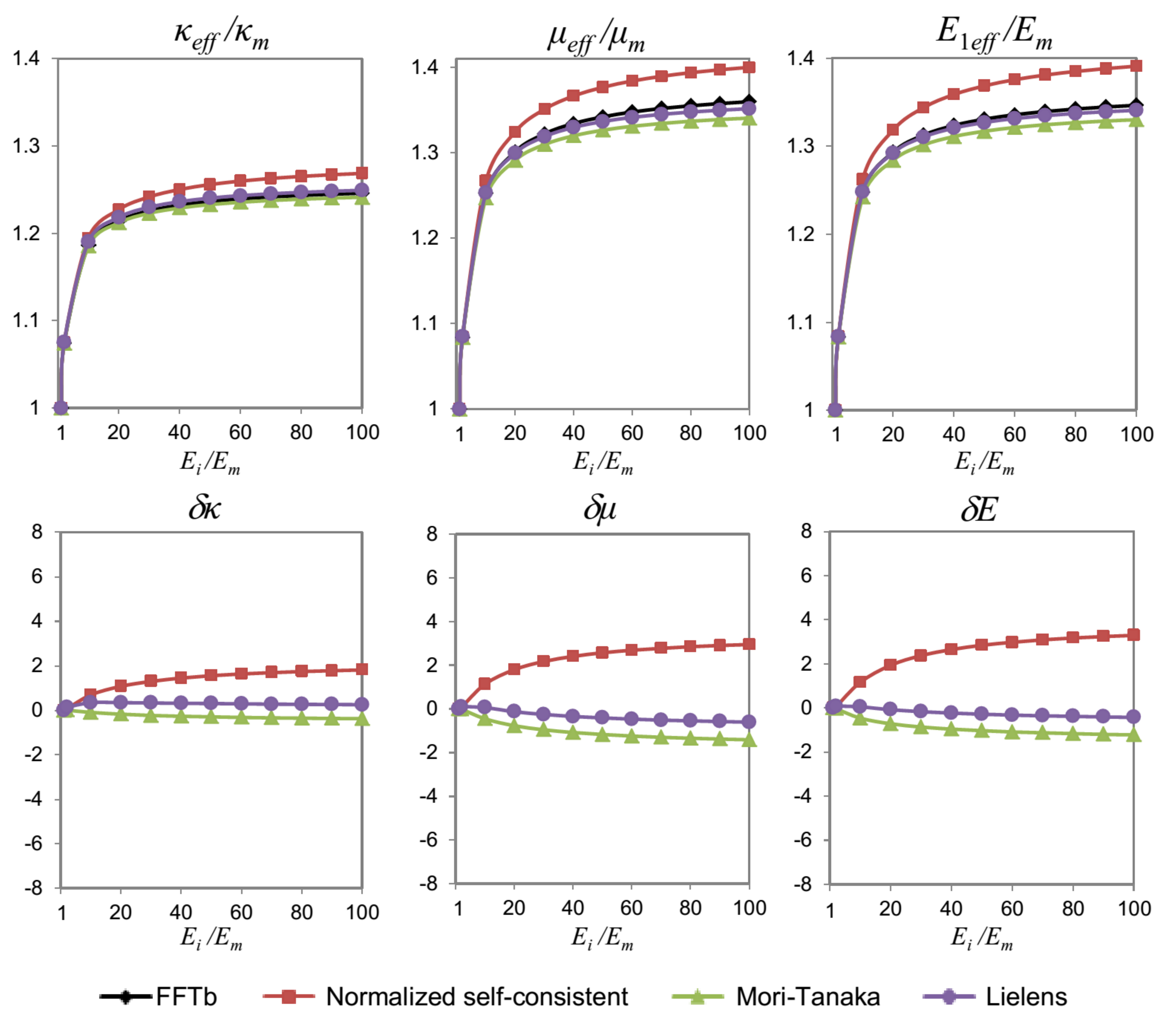}  
 
\caption{\label{fig:results4}Normalized effective bulk, shear and Young's moduli (in direction 1) as function of the Young's modulus contrast for the RVE$_4$ and the corresponding relative deviations (in \%) from the FFTb results. } 
 \end{figure}

 Fig. \ref{fig:results4} shows the normalized effective moduli obtained for RVE$_4$ containing a mixture of spherical and ellipsoidal inclusions in the same proportion (in volume) as RVE$_1$ (5\% of spheres and 6.7\% of ellipsoids). Unlike the RVE$_1$, the ellipsoids were oriented in only 2 directions different from the main directions of the RVE. It is clear that the macroscopic behaviour will not be strictly isotropic in this case. The aspect ratio of the ellipsoids is 5. The models still provide a good estimate for the effective moduli. Lielens and Mori-Tanaka models give, for the volume fractions investigated, the closest predictions, regardless of the macroscopic behaviour is not stricltly isotropic. The normalized self-consistent model is less accurate than the others models but its maximal relative deviation from the FFTb results observed does not exceed 3.5\%. By comparing the effective properties of RVE$_1$ and RVE$_4$ (Figs. \ref{fig:results1} and \ref{fig:results4}, respectively), one can also notice that the mean-field models are slightly sensitive to the ellipsoids aspect ratio because the discrepancies observed are not neglectable for the RVE$_1$ where the ellipsoids exhibits the great aspect ratio.

 Many works were conducted to highlight the validity domains of these mean-field micro-mechanical models (by comparing them to full-field approaches) when predicting the elastic properties of spherical particles %reinforced composites ~\citep{17,18,20}. 
%For example, several spherical particles reinforced composites with various volume fractions of inclusions, up 50\%, were considered in [17]. The authors showed that for inclusion volume fraction of about 10\%, predictions of effective shear and bulk moduli by the mean-field approaches are almost satisfactory. Around 30\% of inclusions, mean-field models are accurate for low contrast and deviate from the FFTb solution for high contrasts. When the inclusions volume fraction reaches 50\%, the same behaviour was observed. Consequently, our study was no more focused on this kind of RVEs. However, simulations were made, confirming these results. For example, Fig. 6 shows the normalized effective shear, bulk and Young’s moduli evolutions as function of the contrast for the RVE$_5$ consisting of 50\% of spherical inclusions. 

 \begin{figure}[!ht]  
\centering
 \includegraphics[width=1\linewidth]{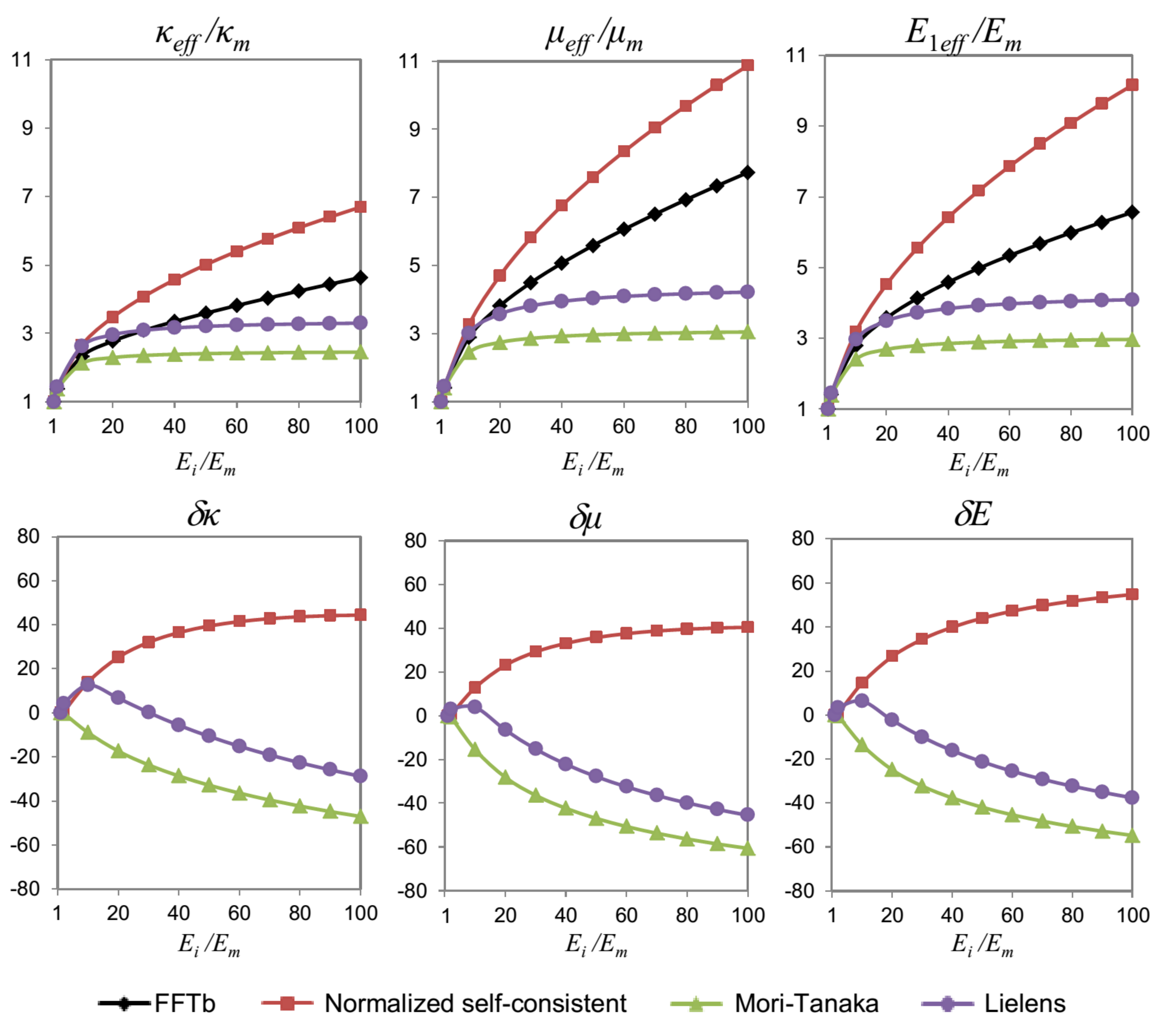}  
 
\caption{\label{fig:results5}Normalized effective bulk, shear and Young's moduli (in direction 1) as function of the Young's modulus contrast for the RVE$_5$ and the corresponding relative deviations (in \%) from the FFTb results.} 
 \end{figure}

 The predictions of Mori-Tanaka and normalized self-consistent models diverge very rapidly when the contrasts increase. The Mori-Tanaka model underestimates the solution obtained with FFTb, while the normalized self-consistent model overestimates it. For low contrasts (up to 50 for the bulk modulus and up to 20 for the shear and Young's moduli), Lielens' model is a very good candidate. Indeed, it leads to a relative deviation from the FFTb model less than 10\%. One can notice, the relative deviations are, in general, more important for the shear and Young's moduli than the bulk one for all the models.

 \subsection{Influence of the inclusion volume fraction on the accuracy of the models}
 
 In this part, the homogenized properties of composites made of randomly oriented isotropic ellipsoidal inclusions distributed into an isotropic matrix were computed for different volume fractions of inclusions. The aim of this section is to highlight the discrepancy  of the mean-field versus full-field approaches while predicting the mechanical behaviour of these kinds of composites when the inclusions’'volume fraction increases. To do this, 6 RVEs containing respectively 4, 8, 12, 16, 20 and 30\% (in volume) of ellipsoidal inclusions have been investigated. Each unit cell contains 10 ellipsoidal inclusions randomly distributed. This distribution (orientation factor) differs from one RVE to another. For this study, the aspect ratio of the ellipsoids is taken equal to 5 for all inclusions.
 
 Using each RVE's parameters given by the RSA and MD-based generation algorithms, we computed the homogenized properties of each RVE. The results obtained in the previous section show that the normalized self-consistent model leads to the largest discrepancy between the three mean-field models investigated in this work, especially when the volume fraction of the inclusions increases. Furthermore, for this model, the computation time increases drastically and the algorithm involves a high memory consumption due to the iterative resolution. For all these reasons, this model was not studied in this part. 
 
\begin{figure}[!ht]  
\centering
 \includegraphics[width=1\linewidth]{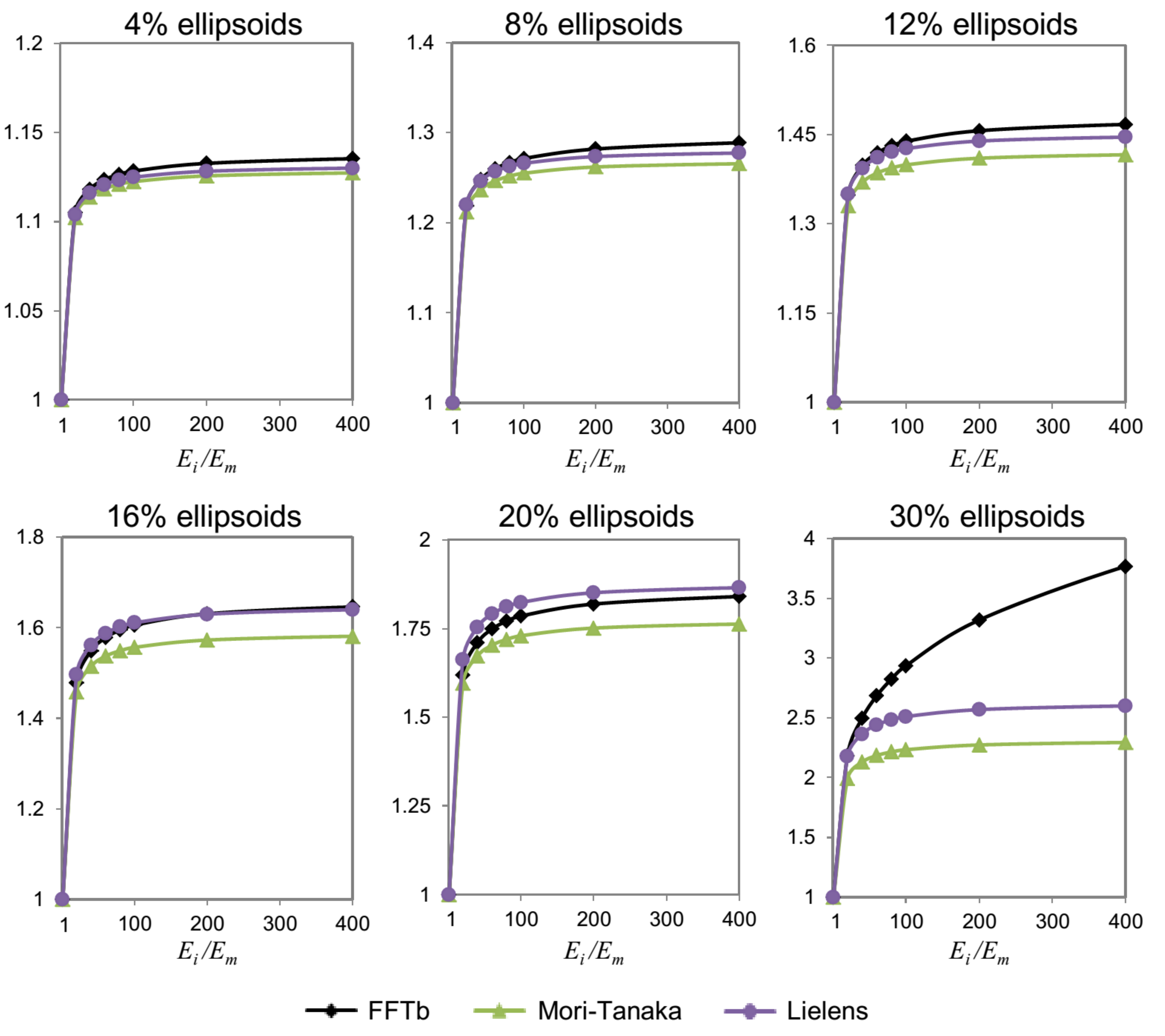}  
 
\caption{\label{fig:results7}Normalized effective shear modulus as function of the Young's modulus contrast for RVEs consisting of ellipsoidal inclusions with a volume fraction of 4, 8, 12, 16, 20 and 30\%, respectively.} 
 \end{figure}

\begin{figure}[!ht]  
\centering
 \includegraphics[width=1\linewidth]{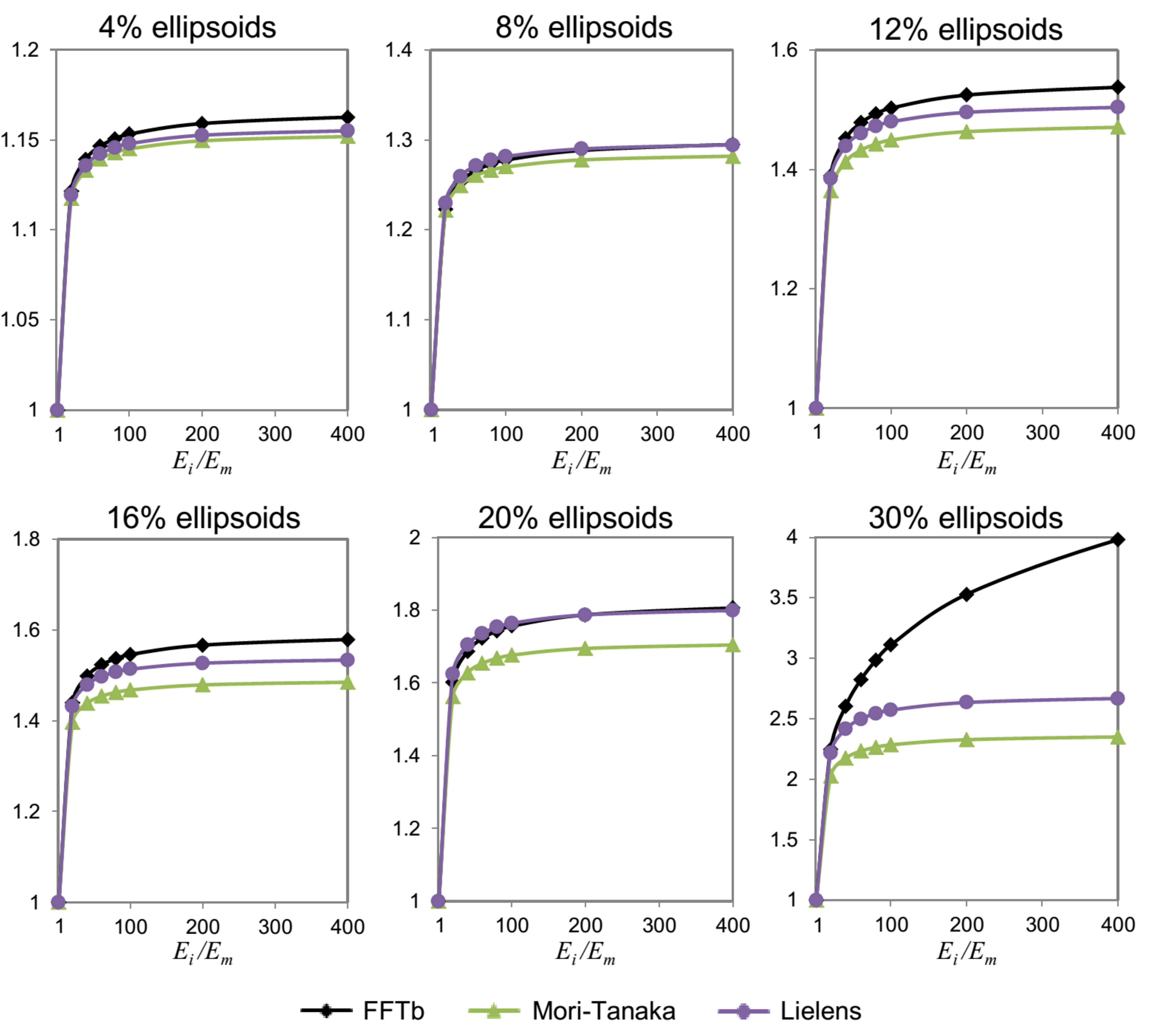}  
 
\caption{\label{fig:results8} Normalized effective Young's modulus (in direction 1) as function of the Young's modulus contrast for RVEs consisting of ellipsoidal inclusions with a volume fraction of 4, 8, 12, 16, 20 and 30\%, respectively.} 
 \end{figure}

\begin{figure}[!ht]  
\centering
 \includegraphics[width=1\linewidth]{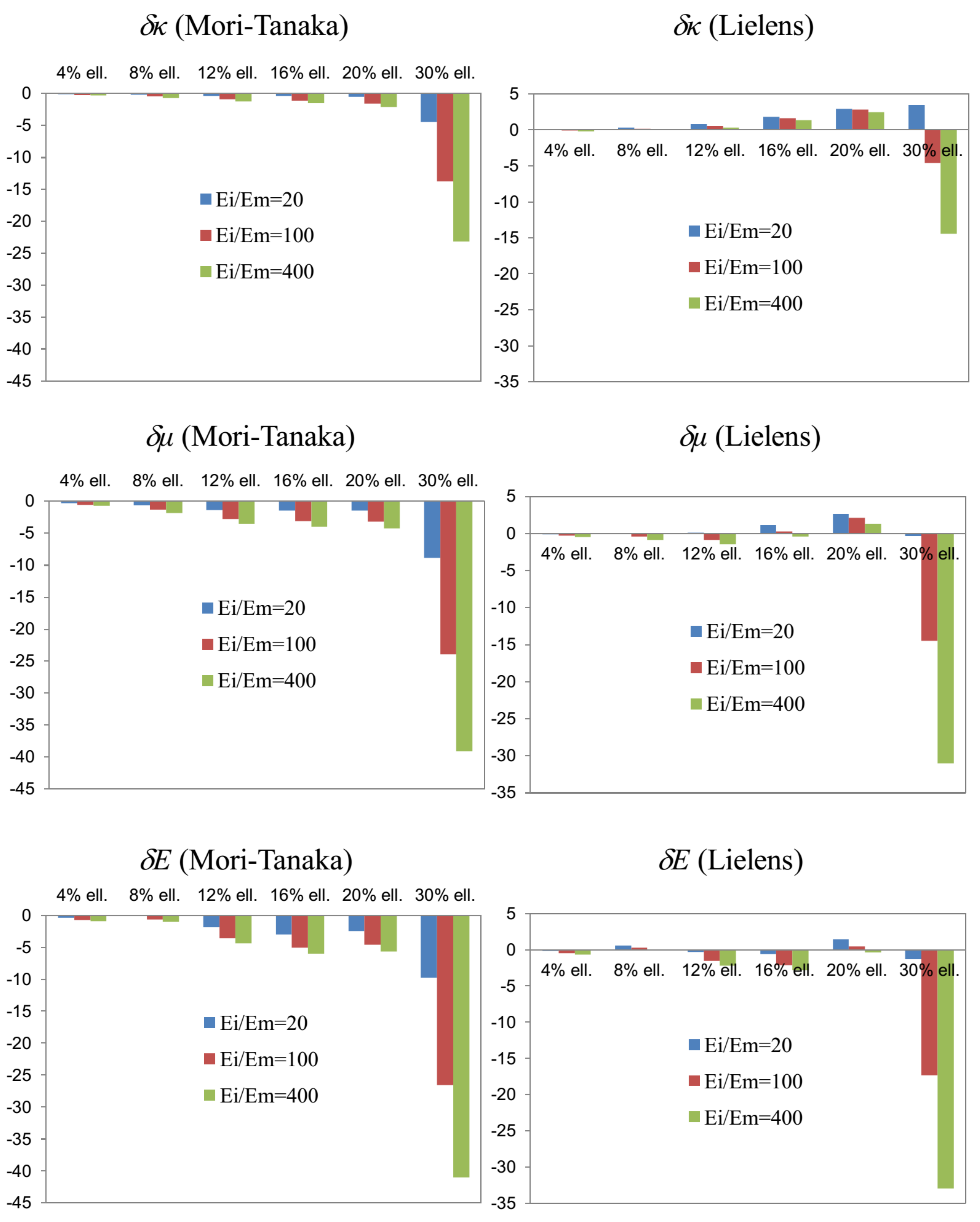}  
 
\caption{\label{fig:results9}Relative deviations (in \%) between Mori-Tanaka and Lielens' models predictions of the effective bulk, shear and Young's moduli and the FFTb results for the contrasts 20, 100 and 400 for the RVEs containing 4, 8, 12, 16, 20 and 30\% of ellipsoidal inclusions. } 
 \end{figure}

 Figs. \ref{fig:results7}, \ref{fig:results8} and \ref{fig:results9} show the evolution of the normalized effective bulk, shear and Young's moduli as function of the Young's modulus contrast for the 6 RVEs investigated. In this study, we extended our computations to the cases when the inclusions are much stiffer. So, the Young's modulus contrast varies from 1 to 400. Note that the RVE$_3$ of the previous section and the one containing 30\% of ellipsoidal inclusions in this section have the same volume fraction of inclusions and the ellipsoids exhibited the same aspect ratio. The difference between these latter is the geometrical orientation of the inclusions. In others words, the orientation factors of the two RVEs are not the same. This explains the slight difference between the results showed in Fig. \ref{fig:results4} for RVE$_3$ and those obtained for the RVE containing 30\% of ellipsoids in Figs. \ref{fig:results7} to \ref{fig:results9}.
By observing the results plotted in Figs \ref{fig:results7} to \ref{fig:results9}, one can notice that Lielens’'model is the most accurate when predicting all the effective properties. The mean-field analytical models are more sensitive to the ellipsoidal inclusions' volume fraction. When the volume fraction of the inclusions increases, the prediction of the analytical models becomes less accurate. Up to about 20\% of inclusions, both Lielens and Mori-Tanaka models deliver accurate estimates of the homogenized properties whatever the contrast. However, the models are more accurate when predicting the effective bulk modulus than the prediction of the other effective properties. From a volume fraction of 30\% of inclusions, the models deliver accurate estimates only for low contrasts: up to 200 for Lielens' model when predicting the normalized effective bulk modulus ($\delta K$ = -9.4\%) and up to only 60 for the prediction of the normalized effective shear ($\delta \mu$ = -9.1\%) and Young's ($\delta E$ = -11.5\%) moduli.

\begin{figure}[!ht]  
\centering
 \includegraphics[width=1\linewidth]{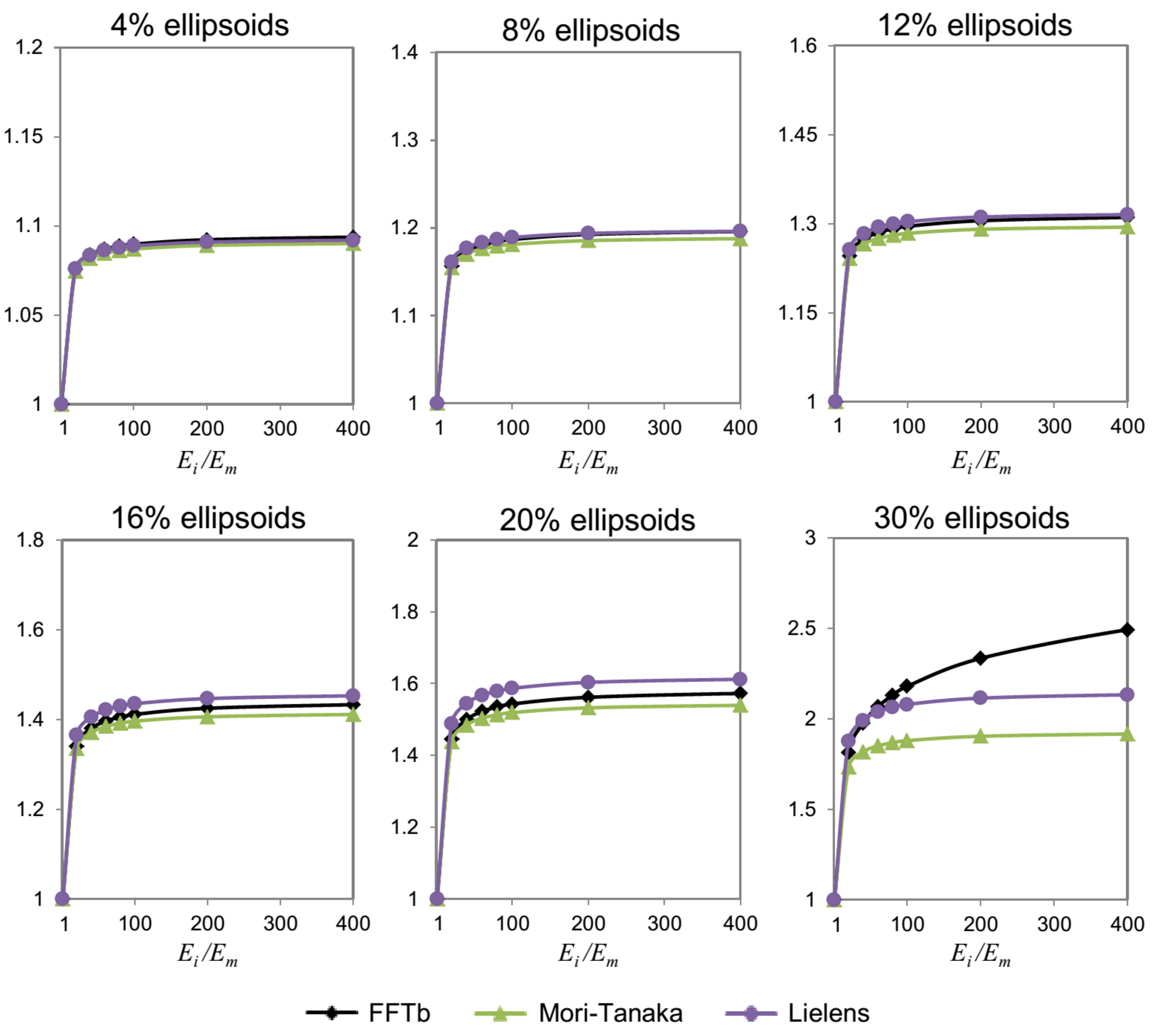}  
 
\caption{\label{fig:results6}Normalized effective bulk modulus as function of the Young's modulus contrast for RVEs consisting of ellipsoidal inclusions with a volume fraction of 4, 8, 12, 16, 20 and 30 \%, respectively. } 
 \end{figure}

 %\subsection{Sensibility to the inclusions geometrical orientation}

\section{Conclusion / outlook}

The influence of the inclusions' morphology on the estimation and the discrepancy between some classical mean-field approximation methods and a full-field computational method based on numerical homogenization techniques to predict the mechanical properties of these materials were investigated in this paper. Several Representative Volume Elements containing spherical, ellipsoidal inclusions and a mixture of  both were studied. For low volume fractions of inclusions, the results of the mean-field approximations and those of the Fast Fourier Transform-based (FFTb) full-field computation are very close, whatever the inclusions morphologys are . In this case, both Lielens' and Mori-Tanaka models can be a good alternative to the FFTb homogenization methods. For RVEs consisting of spherical, ellipsoidal or a mixture of ellipsoidal and spherical inclusions, when the inclusions volume fraction becomes higher, one observes that Lielens' model and the FFTb full-field computation give almost similar estimates. The accuracy of the computational methods depends also on the shape of the inclusions and their volume fraction. The contrasts between the fibers and the matrix remain the most influent parameters on the homogenization models accuracy.The ellipsoids aspect ratio has also some influence on the estimates and the discrepancies between models but this one is lesser than the influence of the volume fraction and the contrast. For microstructures with a mixture of ellipsoidal and spherical inclusions, Lielens' and Mori-Tanaka models could be a good alternative to the FFTb model when the total inclusions volume fraction is about 12\%. In this case, the normalized self-consistent model could also be an alternative to the FFTb model with an error less than 5\%. The normalized self-consistent model is less reliable for matrix-based composite when the inclusion volume fraction is high. Although the full-field computation based on FFT are more sensitive to the contrast  rather than mean-field computation, the homogenization procedure based  on the resolution of the Lippmann-Schwinger equation  with FFT gives a fast, reliable and an efficient way to determine the effective mechanical properties of composites reinforced with ellipsoidal and spherical particles due to less computational time consumption compare with mean-field  homogenization computation.

\section*{Acknowledgements.} Most of the computations described in this paper have been carried out 
at the cluster of the Center of Informatics Resources of Higher Normandy
(CRIANN -- Centre  R\'egional Informatique et d'Applications Num\'erique de Normandie).
Current research of V.S. is supported by the Fonds National de la Recherche, Luxembourg, project F1R-MTH-AFR-080000.

\end{document}